\documentstyle[epsf]{mn}
{}
\begin{document}
\title[Double degenerate close binary]
    {A detached double degenerate with a 1.4 hr orbital period}

\author[C.Moran, T.R.Marsh and A.Bragaglia]
{C. Moran$^{1}$, T. R. Marsh$^{1}$, and A. Bragaglia $^{2}$\\
$^{1}$University of Southampton, Department of Physics, Highfield, Southampton SO17 1BJ. \\
$^{2}$Osservatorio Astronomico di Bologna, via Zamboni 33, I-40126 Bologna, Italy. \\
}

\date{Accepted \\
Received : Version Date of current version \\
In original form 1996 September 19}

\maketitle

\begin{abstract} 

We have discovered that the detached double degenerate binary WD
0957-666 has an orbital period of 1.46 hours, rather than the 1.15 day
orbital period reported earlier. This is the shortest period example
of such a system yet discovered. We obtain a unique period, which fits
both our and earlier data. At this period the emission of
gravitational radiation will cause the binary to merge within
approximately 2.0 x 10$^{8}$ years. This system represents a
population of short orbital period binaries which will merge within a
Hubble time, and so could account for type Ia supernovae, although due
to the low mass of both stars (0.3 to 0.4 M$_{\odot}$), it is unlikely to become a supernova itself. We have
detected the companion star and have measured a mass ratio of q =
1.15$\pm$0.10. This is the third double degenerate for which q has
been measured and all three have q $\simeq$ 1, which is in
conflict with the predicted mass ratio distribution which peaks at
0.7. This system is viewed close to edge on, and we estimate that the
probability of this system undergoing eclipses is 15 \%.

\end{abstract}


\section{Introduction}

One driving force for the detection and study of binary white dwarf,
or double degenerate (DD) systems, is the suggested origin of type Ia
supernovae (SNIa) involving the coalescence of two white dwarfs
(Webbink 1979, Webbink 1984, Iben $\&$ Tutukov 1984 and Branch et al
1995). In this scenario, the binary's orbital period decreases due to
the emission of gravitational radiation, until the less massive white
dwarf is disrupted by tidal forces and is accreted onto its companion
(Mochkovitch and Livio 1989). In order to produce a SNIa, the
coalesced star probably must exceed the Chandrasekhar mass limit of 1.4
M$_{\odot}$, which requires the binary to consist of two CO white
dwarfs. Helium white dwarf binaries have a total mass of 0.5 - 0.75
M$_{\odot}$ (Webbink 1984) and so it is unlikely that a He+He or even
a He+CO binary could contain sufficient mass to produce a SNIa. A
second requirement is that the initial orbital period of the binary is
short enough that merging may occur within a Hubble time. Branch et al
(1995), have considered the different possible progenitors of SNIa, and
favour the coalescence of CO white dwarf pairs. However, they stress
the need for better determination of the properties of double
degenerate systems, and the identification of super-Chandrasekhar mass
systems which will merge within a Hubble time.

Surveys sensitive to short orbital period (P $<$ 3hrs) binaries have
been conducted by Robinson \& Shafter (1987), and Bragaglia et al
(1990). Between them they include a total of 90 DA white dwarfs,
including WD 0957-666, which is the only system in their total sample
to be conclusively proven to be a short orbital period, double
degenerate binary. Foss, Wade, \& Green (1991), conducted a survey
sensitive to orbital periods between 3 and 10 hours, and found no
double degenerates from a sample of 25 white dwarfs. Although the lack
of success was disappointing, it may not be surprising as on current
estimates only 1/360$^{\rm th}$ of the total observable white dwarf
population are type Ia supernovae progenitors (Iben et al 1997)

Double degenerates develop from main sequence binary systems which
undergo 2 separate mass transfer events as each star evolves to fill
its Roche lobe. When mass transfer is dynamically unstable, the
accreting object is unable to accept all the in-falling material, and
a common envelope (CE) is expected to develop. The binary is embedded
in this envelope, into which gravitational drag and tidal forces
deposit orbital angular momentum and energy from the binary. If
sufficient energy is transfered from the binary, the envelope may be
ejected. The efficiency with which the envelope is ejected determines
the extent to which the orbital period of the binary will be
reduced. It is difficult to calculate the orbital period distribution
of double degenerate systems theoretically, and hence to ascertain
whether they could be a viable source of SNIa progenitors, primarily
due to the uncertainty in the efficiency of envelope ejection. This
efficiency is represented by $\alpha_{CE}$, the ratio of the
envelope's binding energy to the orbital energy lost in ejecting the
common envelope. Efficient ejection for which $\alpha_{CE}$ = 1,
results in the orbital period distribution peaking at approximately 12
hours (Yungelson et al 1994). Less efficient ejection will promote
mergers and a orbital period distribution dominated by shorter period
systems, which will be candidates for SNIa progenitors due to their
short merging time scales. The SNIa frequency due to DD mergers
calculated by Han et al (1995) is increased by a factor of 1.8 by
reducing $\alpha_{CE}$ from 1 to 0.3, even though the actual birthrate
of DD systems falls by a factor 2.2 due to early mergers.

As the value of $\alpha_{CE}$ has such a large effect on the physical
parameters of DD systems, observations of these systems can be used to
put tight constraints on the value of $\alpha_{CE}$. With this in mind
we continued our program to detect DDs from a sample of low mass white
dwarfs (Marsh, Dhillon \& Duck 1995, and Marsh 1995), and to measure their
orbital periods, and if possible their mass ratios. 
 
WD 0957-666 was identified as a DD system by the detection of radial
velocity variations resulting from orbital motion. The absence of any
main-sequence star features at long wavelengths implied that the
companion is also degenerate (Bragaglia et al 1990). We observed it in
the hope of being able to detect the companion and hence measure the
mass ratio. The only two measured values of the mass ratio for
detached double degenerates, L870-2 (Saffer et al 1988) \& PG 1101+364
(Marsh 1995), are both around q=0.9, suggesting a possible flaw in the
theoretical models, which peak at 0.7 (Iben et al 1997). Further
measurements of mass ratios and orbital periods for DDs, are crucial
in order to calibrate the theoretical predictions for close binary
evolution.

\begin{figure} 
\begin{center}{
\epsfxsize 0.99\hsize
\leavevmode
\epsffile{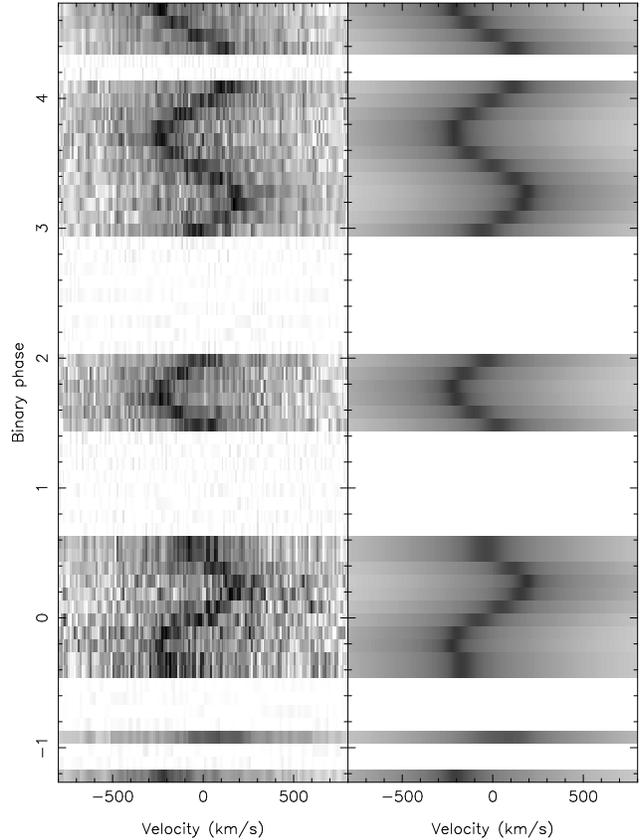}
}\end{center}
\caption{In the left-hand panel are the trailed spectra, showing the large ($\simeq$220 km/s radial velocity shifts of the primary. We show the
model fit to the data in the right-hand panel.}  
\end{figure}  

\section{Observations}

Our data were taken from the 1st to the 3rd of March 1996, with the RGO
spectrograph on the 3.9m Anglo-Australian telescope, with seeing
ranging from 0.6 to 1.2 arc seconds. The spectra cover 6437 to
6677 \AA \  at a dispersion of 0.23 \AA /pixel and a FWHM resolution of
0.7 \AA. By observing H$\alpha$ at this resolution we could resolve the
sharp non-LTE cores of the two stars.

We initially observed WD 0957-666 with 1400 second exposures, but were
surprised when the first two exposures showed a large radial velocity
shift of $\sim$250kms$^{-1}$. Having checked our wavelength scales it
was evident that the orbital period was much shorter than the 1.15
days we had expected. We reduced the subsequent exposures to 500
seconds to minimise the effect of smearing on the spectra as the
binary components shifted wavelength during the exposures. With hind
sight it turns out that the 1400 sec exposures covered $\sim$ 28$\%$
of the orbital period of binary! The shorter exposures represented a
balance between improving the temporal sampling, while maintaining a
reasonable signal-to-noise ratio. Typical 1$\sigma$ errors in the
radial velocity measurements for the 500s exposures are 10
kms$^{-1}$. A total of 33 spectra were taken over the three nights,
including one series which followed a complete orbital cycle.

A narrow (0.8 arcsec) slit was used on all nights to reduce radial
velocity errors. Each group of three to four object spectra were
bracketed by copper-argon arc spectra. The arc and object spectra were
extracted at the same position on the detector, and the wavelength
scales for the object spectra were derived by interpolating the
surrounding pair of arc spectra with time.

\begin{figure} 
\begin{center}{
\epsfxsize 0.99\hsize
\leavevmode
\epsffile{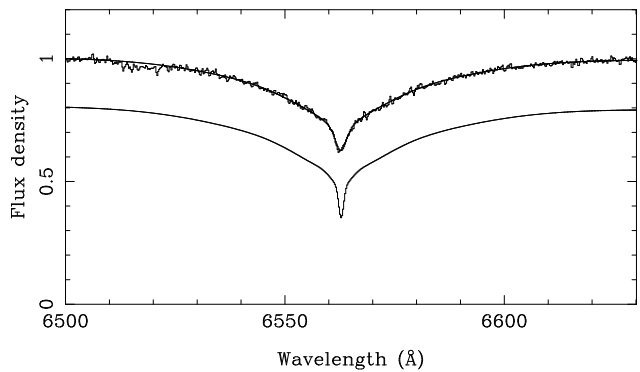}
}\end{center}
\caption{At the top we show the mean spectrum of WD 0957-666 corrected
for orbital motion, along with the model fit. Telluric absorption
can be seen around 6520 \AA. Off-set below this is the true
model profile, unaffected by the smearing caused by orbital motion
during the exposures.}  
\end{figure}

\begin{figure*} 
\begin{center}{
\epsfxsize 0.99\hsize
\leavevmode
\epsffile{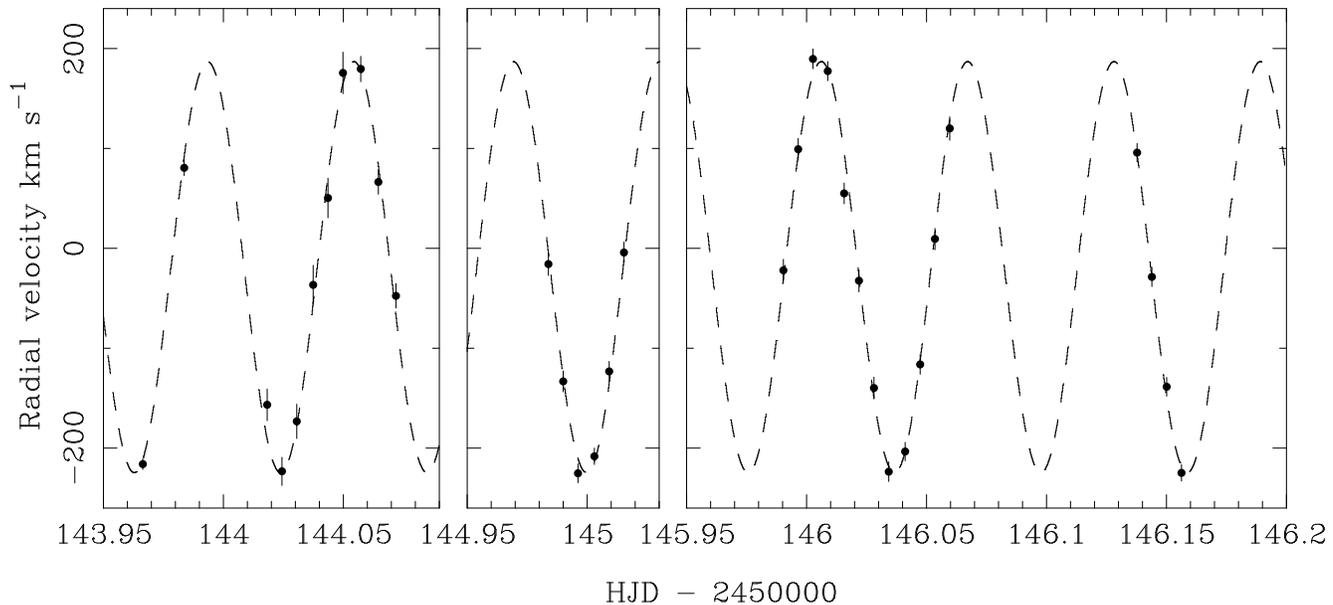}
}\end{center}
\caption{The measured radial velocities for each of our 33 spectra,
along with the circular orbit fit.}  
\end{figure*} 

\section{Results}

The raw data and the model fits are shown together in figure 1. The
measurement of radial velocities followed the procedure described in
Marsh et al (1995), but with some adaptations to account for the
smearing of the spectra caused by the rapid orbital motion. The
spectra were normalised and then averaged. The average was fitted with
a model consisting of a straight line and multiple Gaussian
components, fixed to have the same velocity for any given
spectrum. The FWHMs and heights of the Gaussians were then held fixed
while their velocities were allowed to vary. The resultant velocities
were fitted with a circular orbit. The velocities from this
orbital fit were then removed from the spectra which were then
re-averaged (Fig 2).  The cycle of averaging, fitting, removing
velocities and then re-averaging was repeated three times. As a result
of cycling the fitting process, the fit sharpened considerably,
requiring the introduction of more Gaussian components. Four Gaussians
were used in the final fit, and the radial velocity measurements
converged to stable values, with a semi-amplitude velocity of K$_{1}$
= 196 km s$^{-1}$, see figure 3.

To refine the orbital period, we re-measured radial velocities of
H$_{\beta}$ in 24 spectra taken between Jan 1988 and Jan 1990,
(Bragaglia et al 1990). The Scargle periodogram in figure 4 shows how
the extension of the observational baseline can increase the accuracy
with which the orbital period can be determined. The combined data set
of measured radial velocities was fitted with periods derived from
many of the most likely orbital period aliases shown in the inset of
figure 4. The $\chi^{2}$ values for the orbital fits rose sharply for
periods other than that represented by the peak alias in the
periodogram, ruling out any other orbital periods.

At this stage we adapted the fitting routine to include the effects of
smearing due to orbital motion. The exposure time and orbital period were used
to calculate the phase width covered by each spectrum. Each model fit
was then calculated by trapezoidal integration, accounting for the
orbital motion during the exposure.  The inclusion of the smearing
parameter had the expected effect of further sharpening the fits, and
of increasing the semi-amplitude of the circular orbital fit to
K$_{1}$ = 205 km s$^{-1}$. The fitting cycle was repeated three more
times with smearing accounted for, until the radial velocity
measurements converged on stable values. The circular orbit fit to our
and Bragaglia et al's data is shown in figure 5.

\begin{figure} 
\begin{center}{
\epsfxsize 0.99\hsize
\leavevmode
\epsffile{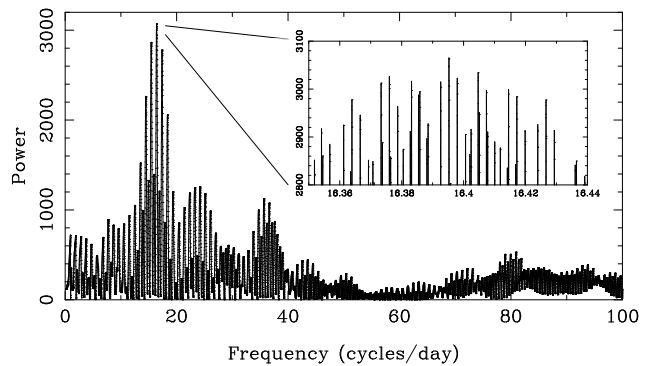}
}\end{center}
\caption{A periodogram of our data shows the dominance of a 16
cycles/day orbit. The extension of the observational baseline with the
inclusion of Bragaglia et al's data allows the different orbital aliases to be
resolved, as shown in the inset.}  
\end{figure} 

\begin{figure} 
\begin{center}{
\epsfxsize 0.99\hsize
\leavevmode
\epsffile{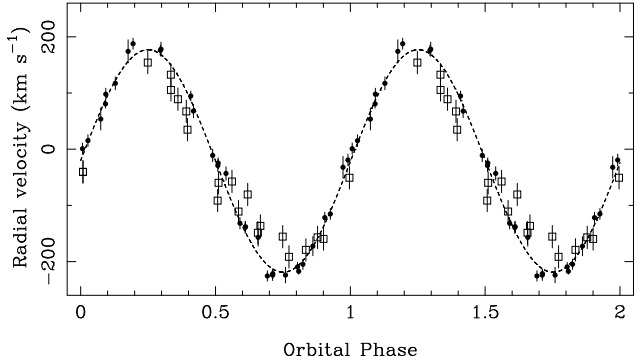}
}\end{center}
\caption{The phase folded radial velocity measurements and circular
orbit fit for WD 0957-666. The solid circles represent our data while
the squares represent measurements from Bragaglia's data.}  
\end{figure}  

\subsection{Detection of the Companion}

So far, only the primary star had been modelled, as there was no
obvious sign of the secondary star, in either individual spectra, or the
trailed spectra (Fig 1). Detecting the companion is highly desirable as it
allows calculation of the mass and luminosity ratios of the binary. 

\begin{figure*}
\begin{center}{
\epsfxsize 0.99\hsize
\leavevmode
\epsffile{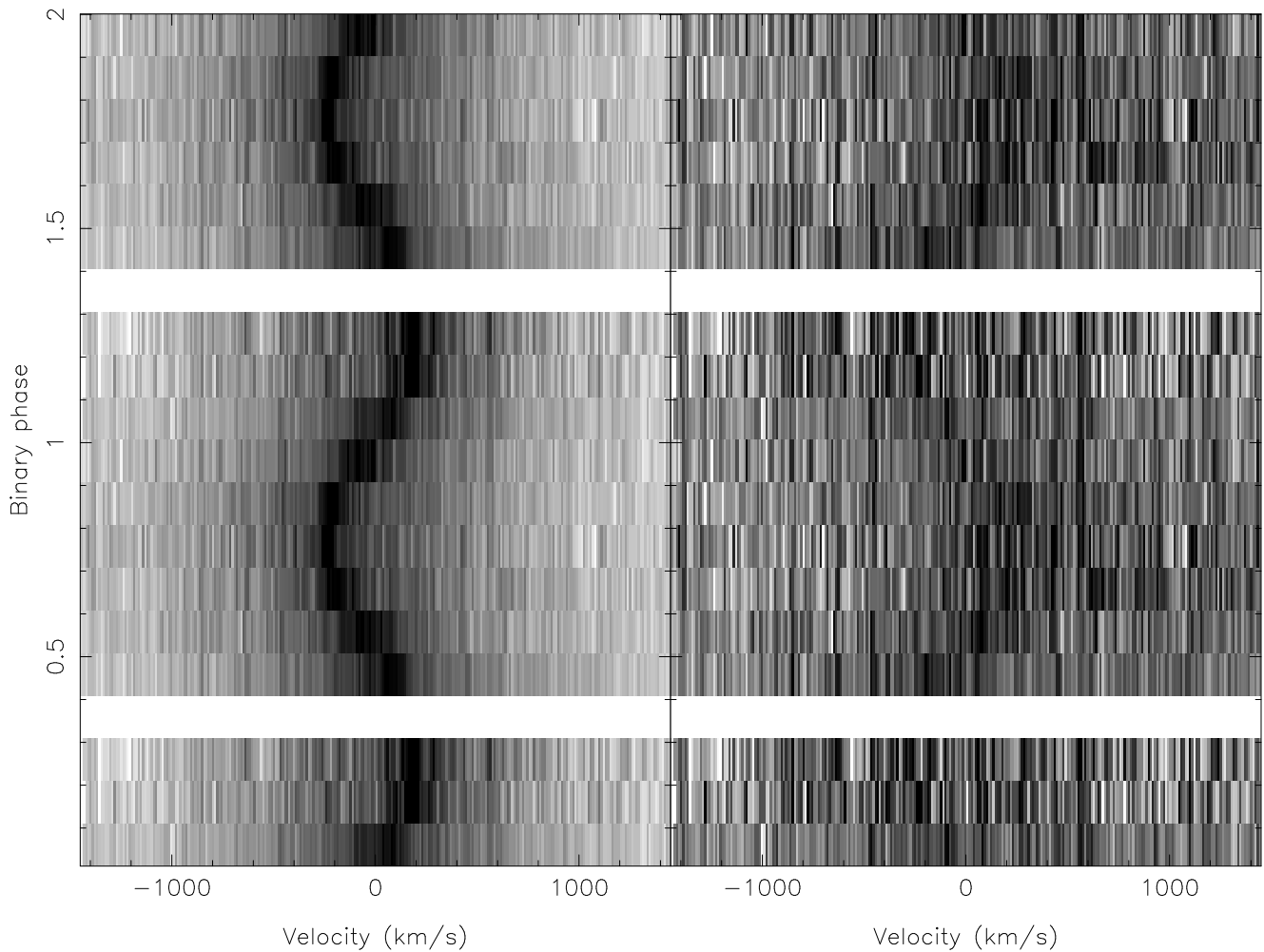}
}\end{center}
\caption{In the left hand panel we show the phase binned raw spectra,
which shows the dominant primary star. The right hand panel contains the
residual from subtracting the model fit of the primary star, from the
data. The faint secondary star is detected with a semi-amplitude of
approximately 250 km/sec, in anti-phase to the primary star. Some weak
telluric features appear as vertical lines in this panel.}  
\end{figure*} 

We subtracted the model fits from the data and then phase
binned the residuals to increase the S/N. We show the results in
figure 6. Although very faint, the companion has been detected. As the
companion appears in absorption it confirms the double degenerate
nature of WD 0957-666. When measured by eye, the companion appears to
have a semi-amplitude K$_{2}$, of approximately 200 km s$^{-1}$, and
appears to be 180$^{\circ}$ out of phase with the primary as expected.

\begin{table}
 \centering
 \caption{Model fitting parameters}
 \begin{tabular}{ccccc}
Constant & Gaussian & Star & FWHM(\AA) & Height (mJy)\\
1.017 & 1 & 1 & 53.4 & -0.150 \\
      & 2 & 1 & 20.2 & -0.109 \\
      & 3 & 1 & 4.91 & -0.062 \\
      & 4 & 1 & 1.30 & -0.127 \\
      &   &   &      &        \\
      & 1 & 2 & 40.7 & -0.041 \\
      & 2 & 2 & 4.97 & -0.023 \\
\end{tabular}
\end{table}

\subsection{Orbit fitting to the companion} 

Having detect the companion, we determined its radial velocity
semi-amplitude by including two further Gaussian components to the
previously used model fit, to represent the secondary star. As no
individual spectrum shows any strong evidence of the companion, we
applied the fit to all the spectra, fitting FWHM, heights and orbital
parameters simultaneously. The final gaussian model fitting parameters
are shown in table 1. The fitting procedure converged to give stable
values for the companion's orbital parameters, as shown in table
2. Our best fit value of K$_{2}$ was 252.4$\pm$21.5 km s$^{-1}$. The
inclusion of the Gaussian components to model the companion star,
increased the radial velocity semi-amplitude of the primary K$_{1}$ to
219.8$\pm$1.9 km s$^{-1}$, as the fit to the primary star was no longer
pulled to lower velocities by the presence of the unaccounted for
secondary star.

\begin{table}
 \centering
 \caption{Circular orbit fit parameters}
 \begin{tabular}{ll}
$\chi^{2}/D of F$ &0.969 \\
N & 33 \\
$\gamma$ (km s$^{-1}$) & -18.7 $\pm$ 1.4 \\
K$_{1}$(km s$^{-1}$) &  219.8 $\pm$ 1.9 \\
K$_{2}$(km s$^{-1}$) & 252.4 $\pm$ 21.5 \\
Orbital Period (d) & 0.0609931806 $\pm$ 8.4$^{-10}$ \\
T$_{o}$ & 2450145.3796 $\pm$ 1$^{-4}$ \\
\end{tabular}
\end{table}

\section{Discussion}
\subsection{The mass of the companion}

The mass ratio, as defined below, may be calculated directly using the
semi-amplitude velocities found with the fitting process. Using
K$_{1}$ = 220$\pm$2 km s$^{-1}$ and K$_{2}$ = 252$\pm$21 km s$^{-1}$
we calculate q = 1.15$\pm$0.10. 

\[ q = \frac{M_{1}}{M_{2}} = \frac{K_{2}}{K_{1}} \]

Bragaglia et al (1995) determined the surface gravity and effective
temperature of the primary star to be log(g) = 7.285$\pm$0.082 and
\(T_{eff} = 27047\pm398 K\) respectively. They calculated the primary
star mass to be \(M_{1} = 0.335\pm0.018 M_{\odot}\) using models with
carbon-oxygen cores. Bergeron et al (1992) have shown that the zero
temperature mass-radius relationships of Hamada \& Salpeter (1961) are not
significantly different for carbon and helium configurations at this
mass. However, using new evolutionary models with helium core
configurations (Althaus \& Benvenuto 1997) that include finite
temperature effects (which are most significant for low mass stars) we
estimate that \(M_{1} = 0.37\pm0.02 M_{\odot}\). This leads to a mass
determination for the secondary star of M$_{2}$ = 0.32$\pm$0.03
M$_{\odot}$. 

Although we cannot calculate the individual masses due to the unknown
inclination of the system, $i$, we can calculate the following quantities.

\[ M_{1}\sin^{3}i = \frac{P}{2\pi G}(K_{1}+K_{2})^2K_{2} = 0.356\pm0.066 M_{\odot} \]

\[ M_{2}\sin^{3}i = \frac{P}{2\pi G}(K_{1}+K_{2})^2K_{1} = 0.310\pm0.033 M_{\odot} \]

For the above value of \(M_{1}\sin^{3}i\) to be consistent with the
mass \(M_{1} = 0.37\pm0.02 M_{\odot}\), sin$i$ $\simeq$ 1, and so the
system must be observed at a high inclination. A calculation yields
sin$i$ = 0.99$\pm$0.11. The fact that the system is viewed close to
edge on, raises the tantalizing possibility that this system may
undergo eclipses.

\subsection{Mass ratios of double degenerates} 

The mass ratio estimate is interesting as it implies that the system
could have previously been through an Algol-like phase of conservative
mass transfer (Sarna et al 1996). A value of q $>$ 1 means the brightest
white dwarf is also the most massive. It must then be at a higher
temperature, and hence younger than its companion to be so
visible. The apparent paradox is that the mass of each white dwarf is
effectively the mass of the degenerate helium core of its progenitor
star at the point when it fills its Roche lobe, which depends on the
size of the semi-major axis of the binary at that time. Roche lobe
overflow from a giant with deep convective layers usually leads to a
common envelope phase, resulting in orbit shrinkage as the envelope is
ejected at the expense of orbital energy. Hence by the time the
secondary star (the progenitor of the primary WD) evolves to fill its
Roche lobe, the orbital separation should be significantly smaller and
hence so should the degenerate core of the giant, and the resulting
white dwarf mass. For the youngest white dwarf to be also the most
massive, the system must have undergone a period of conservative mass
transfer, during which the initial mass ratio was reversed so that the
most evolved star became the least massive, and produced the least
massive white dwarf.

\begin{figure} 
\begin{center}{
\epsfxsize 0.99\hsize
\leavevmode
\epsffile{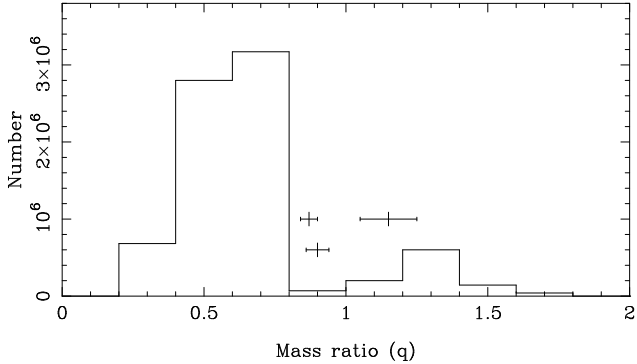}
}\end{center}
\caption{The theoretical mass ratio distribution for close double
degenerate binaries with helium white dwarfs. (Iben et al 1997) The tick marks show the only three measured mass ratios and
their associated errors.}  
\end{figure}  

There are only two other DD systems which have measured mass ratios,
L870-2 for which q = 0.90$\pm$0.04 (Saffer et al 1988) and PG 1101+364
for which q = 0.87$\pm$0.03 (Marsh 1995). This is unexpected as
theoretical predictions suggest a peak at q = 0.7, see figure
7. Having three out of three measured values in the wing of the
distribution profile, suggests a flaw in the theoretical models. A
more accurate determination of q for this, and other DD systems is
needed to confirm whether this tendency for the mass ratio to be close
to 1 is just coincidental, or whether the theoretical predictions are
inaccurate.

\subsection{The possibility of eclipses}

We conducted Monte Carlo simulations to investigate the probability of
eclipses in this binary. We took log T$_{eff}$ = 4.432 and log g =
7.285 from Bragaglia et al (1995), and interpolated between values
given in table 6 of Althaus and Benvenuto (1997), to give the mass
radius relationship. We estimate that the probability of this binary
system undergoing eclipses is 15 \%. The significance of an eclipsing
double degenerate system is that it could be used to check the
accuracy of the equation for the emission of gravitational radiation,
as well as the mass-radius relationship for white dwarfs.

\subsection{The luminosity ratio}

We can estimate the brightness of the companion by assuming that its
line profile has the same shape as that of the primary star. The FWHMs
of the Gaussian components, and their relative strengths, were fixed
to be equal for both stars, while the overall strength of each star
was allowed to vary. These fits were then applied to the data, having
fixed the orbits of both stars. We estimate that the luminosity ratio,
L$_{1}/L_{2} \simeq 5.1$. The assumption of similar profile shapes for
both stars is an obvious simplification, as the companion is much
cooler than the primary star, and so might have a different
profile. In principle this could be accounted for, but is beyond the
scope of this investigation.

\subsection{The evolution of WD 0957-666, past, present and future}

The previous evolutionary pathway of this system is constrained to
some extent by the mass ratio measurements. It seems probable that the
progenitors of these white dwarfs underwent an Algol-like phase in the
past, resulting in the reversal of the mass ratio.  The time passed
since the binary emerged from the last common envelope phase can be
estimated from the primary's temperature T$_{eff}$ = 27047 K
(Bragaglia et al 1995). We estimate that the system has been in its present
state for 10$^{7}$yrs (Marsh et al 1995). This is only a fraction of
the merging time scale (see below) and hence the orbital period of this system
will not have changed appreciably since its formation.

The only changes the system will now be undergoing are the continual
cooling of both white dwarfs, and the reduction of the orbital
separation and orbital period due to the emission of gravitational
radiation. Given the already short orbital period of this system, the
stars will merge rapidly, even considering their low masses. The time
required to merge due to gravitational radiation emission is given by
\[\tau_{m} = 1.00 \times 10^{7}
\frac{(M_{1}+M_{2})^{\frac{1}{3}}}{M_{1}M_{2}}P^{\frac{8}{3}}{\rm
yr}\] with the orbital period given in hours, and the masses in solar
units. Taking \(M_{1} =
0.37 M_{\odot}\), \(M_{2} = 0.32M_{\odot}\) and P = 1.46 hrs, we
calculate that the binary will merge within 2.0 $\times
10^{8}$yrs. WD 0957-666 is a system which will merge well
within one Hubble time.

The future evolution of this system is dependent on the outcome of the
merging process. Given the limits we have placed on the mass ratio of
this system, it is not possible that the companion could be massive
enough for the merged object to exceed the Chandresekar mass
limit. Hence it is unlikely to produce a SNIa, although the existance
of sub-Chandresekar mass progenitors has not been ruled out (Livne \&
Arnett 1995). The most likely outcome will be the formation of a
helium subdwarf star (Webbink 1984).

\subsection{Short orbital period DDs as SNIa progenitors}
 
The detection of this short orbital period system, offers more
observational proof that these systems can be born with very short (P
$<$ 4hrs) orbital periods, as predicted theoretically. A population of
such systems will need to exist if merging white dwarfs are SNIa
progenitors. Also, by populating the short and long period wings of
the theoretical orbital period distribution, we can improve the
estimated value for $\alpha_{CE}$. Choosing different values of
$\alpha_{CE}$ shifts the relative position of the theoretical orbital
period distribution, with an increase in $\alpha_{CE}$ shifting the
distribution to longer orbital periods. Improving theoretical
predictions, and the parameters like $\alpha_{CE}$ on which they are
based, will help to determine whether DDs exist in the rights numbers,
and with the right periods and masses to be the progenitors of SNIa.

\section{Conclusions}
 
We have determined the orbital period for the double lined, detached
double degenerate white dwarf WD 0957-666. At 1.46 hours its the
shortest period system of its kind yet found. The emission of
gravitational radiation will force the binary to merge within only 2.0
x 10$^{8}$ years, but the combined mass will not be sufficient to make
this system a SNIa progenitor candidate. We have measured the mass
ratio of the binary to be q = 1.15$\pm$0.10. All three such systems for
which q has been measured, have a value close to 1. It suggests that
the theoretical mass ratio distribution which peaks at q = 0.7, may be
in trouble. The measured semi-amplitudes of the two stars, along with
the spectroscopic mass of the primary, imply that the system is viewed
at an inclination close to 90$^{\circ}$, with a 15\% probability that
the binary is eclipsing.

\subsection*{Acknowledgments}

This work was based on observations taken from the Anglo-Australian
Telescope, and we would like to thank the staff there for their
help. C. Moran was supported by a PPARC studentship. TRM was supported
by a PPARC Advanced Fellowship.

\bsp

\end{document}